\documentclass[a4paper, 11pt,onecolumn,final,notitlepage,oneside]{article}
\usepackage[cp1251]{inputenc}
\usepackage[T1,T2A]{fontenc}
\usepackage[english]{babel}
\usepackage{indentfirst}%для отступов
\usepackage{geometry}
\usepackage{graphicx}%для включения PS файлов
\usepackage[final]{epsfig}
\usepackage{multicol}%для организации многоколоночного текста (предм. указатель)
\usepackage{graphicx}
\usepackage{subfigure}
\usepackage{amsmath}
\usepackage{enumerate}
\usepackage{amssymb}
\usepackage{amscd}
\usepackage{hhline}
\usepackage{multirow}
\usepackage{setspace}
\usepackage{makeidx}
\usepackage{textcomp}
\usepackage{amsfonts}
\usepackage{afterpage}% Полезно для полного заполнения страниц перед большими таблицами
\usepackage{longtable}%Для таблиц на нескольких страницах
\usepackage{cite}
\usepackage{rawfonts}
\usepackage{array}
\usepackage{amsopn}
\usepackage{blindtext}
\usepackage{hyperref}
\hypersetup{colorlinks=true, linkcolor=blue, citecolor=green, filecolor=magenta, urlcolor=magenta, pdfpagemode=FullScreen}

 \graphicspath{{pictures/}}

\DeclareGraphicsExtensions{.pdf,.png,.jpg,.eps}

\numberwithin{equation}{subsubsection}

\makeatletter
\def\@seccntformat#1{\csname the#1\endcsname.\ } % точка после номера раздела
\def\@biblabel#1{#1.} % формат номеров в списке литературы
\makeatother

\begin{document}

\title{\normalsize \begin{flushleft}
{UDC 531.314.2, 531.395, 531.62, 531-1, 531.011}
\end{flushleft}
\vspace{\baselineskip}
 \normalsize \bf Trajectory equations for a non-conservative natural system}
\author{\bf \small V. V. Voytik\,\\
\small \itshape Department of Medical Physics and Informatics, \\ \small \itshape Bashkir State Medical University,  \\ \small \itshape Lenin St., 3, Ufa, 450008, Russian Federation\\
\small \itshape e-mail: vvvojtik@bashgmu.ru\\}
\date{}
\maketitle
\renewcommand{\abstractname}{}

\begin{abstract}
The purpose of the article is to derive equations that determine the trajectory of a non-conservative natural system in configuration space in non-stationary external fields. A theorem on the change in the kinetic energy of the system is preliminarily proved. Lagrange equations are used to derive the equations. The derived equations can be solved numerically by the Runge-Kutta method of the 4th order. The trajectory equations together with the equality describing its parameterization form the trajectory method for solving dynamics problems.
\end{abstract}

\begin{flushleft}
{\bf{Introduction}}
\end{flushleft}

The mechanical state of a natural system (i.e. its position and state of motion) in Lagrange's formulation is completely determined by setting its initial coordinates and initial velocities. Then the solution of Lagrange's equations of motion leads to the establishment of coordinates as a function of time. But there is also a geometric point of view on mechanical motion. Another, equivalent method consists of determining the initial coordinates of the system in the configuration space and the tangent vector to its trajectory. Then, knowing the kinetic energy of the system as a function of coordinates and time, it is possible to determine the trajectory and parametrize it: i.e., to associate each of its small sections with a certain moment in time.

If a natural system is conservative, then its trajectory equations are in general known. In ray optics they are known as the eikonal equation \cite[equations (1.1.7), (1.1.15)]{19}. For a material point in a constant external non-magnetic field, these equations are given in \cite[problem to p. 44]{1}, \cite[p. 2.7.]{2}). From the point of view of differential geometry, they reduce to the usual second Newton's law, expressed in projection onto the normal vector to the trajectory. Trajectory equations are an essential part of some sections of physics. For example, in geometric optics \cite{3} the equations describing the behavior of monochromatic light rays are considered. In electron optics \cite[chapter 3]{4} the Greenberg equations are considered, describing the theory of focusing a beam of charged particles in constant electric and magnetic fields. The geometric representation of mechanics in the configuration space was also considered in \cite{5}-\cite{22}. In \cite{23} the theorem on the completeness of vector fields in Riemannian Hilbert manifolds for trajectories accelerated by time-dependent forces was discussed.

The purpose of the article is to establish the equations that determine the trajectory of a natural system in the configuration space in the most general case, that is, for non-conservative systems. This will allow us to consider such systems not only by the well-known Lagrange, Hamilton, and Hamilton-Jacobi-Ostrogradsky methods, but also from a visual geometric point of view. Therefore, the equations that describe the trajectory of a natural system in variable fields and with variable energy are important and interesting. In addition to possible far-reaching theoretical consequences, such equations for natural systems, due to their widespread use, can be useful for modeling many devices, for example, of the electron-optical type. The trajectory equations can also be used in various engineering applications: from robotics and aerospace applications to the development of optimal control and navigation algorithms.

\subsubsection{Equation of motion of a natural system in Lagrange form}

 The Lagrangian of a natural system, as is known \cite{8}, \cite{9}, is a quadratic polynomial in the generalized velocities  $\dot{q}^{\alpha}$
\begin{equation}\label{voy1.1}
	L=\frac{1}{2}\mu_{\alpha\beta}\dot{q}^{\alpha}\dot{q}^{\beta}+P_{\alpha}\dot{q}^{\alpha}-U\,.
\end{equation}
The structure of the Lagrange function \eqref{voy1.1} is possessed by an ordinary mechanical system (for example, a deformable body with the moment of inertia $\mu_{\alpha\beta}=\mu_{\alpha\beta}(q^1,q^2,…,q^s,t)$ or a point particle in Riemannian space) possessing potential energy $U=U(q^1,q^2,…,q^s,t)$ and potential momentum $P_{\alpha}=P_{\alpha}(q^1,q^2,…,q^s,t)$ (the term belongs to C. Kittel \cite[Appendix I]{10}) in some electric and magnetic fields.

The momentum along the $q_{\gamma}$ coordinate is equal to 
\begin{equation}\label{voy1.2}
	p_{\gamma}=\frac{\partial L}{\partial \dot{q}^{\gamma}}=\mu_{\gamma\beta}\dot{q}^{\beta}+P_{\gamma}\,.
\end{equation}
The equations of motion in Lagrange form for \eqref{voy1.1} are
\begin{equation}\label{voy1.3}
	\frac{d}{dt}\left(\mu_{\gamma\beta} \dot{q}^{\beta}+P_{\gamma}\right)=\frac{1}{2}\frac{\partial \mu_{\alpha\beta}}{\partial q^{\gamma}}\dot{q}^{\alpha}\dot{q}^{\beta}+\frac{\partial P_{\alpha}}{\partial q^{\gamma}}\dot{q}^{\alpha}-\frac{\partial U}{\partial q^{\gamma}}\,.
\end{equation}

Let us represent the Lagrange equations in a solved form, with respect to covariant accelerations. The total derivative $d{\mu_{\alpha\beta}}/dt$ consists of two parts: the change in the metric tensor over time at a given point in the configuration space and the change at a given moment of time when moving to another point
\begin{equation}\label{voy1.4}
	\frac{d\mu_{\gamma\beta}}{dt}=\frac{\partial\mu_{\gamma\beta}}{\partial t}+\frac{\partial\mu_{\gamma\beta}}{\partial q^{\alpha}}\,\,\dot{q}^{\alpha}\,.
\end{equation}
A similar formula is also valid
\begin{equation}\label{voy1.5}
	\frac{dP_{\gamma}}{dt}=\frac{\partial P_{\gamma}}{\partial t}+\frac{\partial P_{\gamma}}{\partial q^{\alpha}}\,\,\dot{q}^{\alpha}\,.
\end{equation}

Differentiating the left side of \eqref{voy1.3}, taking into account \eqref{voy1.4}, \eqref{voy1.5} and grouping the terms by powers of velocity, we obtain that
	\[\mu_{\gamma\beta}\ddot{q}^{\beta}=-\frac{\partial U}{\partial q^{\gamma}}-\frac{\partial P_{\gamma}}{\partial t}+\left(\frac{\partial P_{\alpha}}{\partial q^{\gamma}}-\frac{\partial P_{\gamma}}{\partial q^{\alpha}}-\frac{\partial \mu_{\gamma\alpha}}{\partial t}\right)\dot{q}^{\alpha}+\]
	\begin{equation}\label{voy1.6}
	+\frac{1}{2}\left(\frac{\partial\mu_{\alpha\beta}}{\partial q^{\gamma}}-\frac{\partial\mu_{\gamma\beta}}{\partial q^{\alpha}}-\frac{\partial\mu_{\gamma\alpha}}{\partial q^{\beta}}\right)\dot{q}^{\alpha}\dot{q}^{\beta}\,.
\end{equation}
These are the Lagrangian equations of motion.

Taking into account the equality \eqref{voy1.2}, the energy of the natural system $E$ as a function of coordinates, velocities and time is equal to
\[	E=p_{\alpha}\dot{q}^{\alpha}-L=\frac{1}{2}\mu_{\alpha\beta}\dot{q}^{\alpha}\dot{q}^{\beta}+U\,.\]
Hence the kinetic energy $T$ is equal to
\begin{equation}\label{voy1.7}
	T=\frac{1}{2}\mu_{\alpha\beta}\dot{q}^{\alpha}\dot{q}^{\beta}=E-U\,.
\end{equation}

The kinetic energy of a natural system, as a known quantity standing in the middle part of the equality \eqref{voy1.7}, can also be understood as an already known function of coordinates and time $E-U$, which stands on the right. Then, expressing from the equality \eqref{voy1.7} $dt$, we obtain
\begin{equation}\label{voy1.8}
	dt=\sqrt{\frac{\mu_{\alpha\beta}dq^\alpha dq^\beta}{2(E-U)}}=\frac{dq}{\sqrt{2(E-U)}}\,,
\end{equation}
where
\[	dq^2=\mu_{\alpha\beta}dq^\alpha dq^\beta\]
is an element of length between close points in the configuration space.

\subsubsection{Statement of the problem}

The sought equations must be related to known methods of mechanics. It turns out that from the point of view of Lagrangian mechanics, the trajectory equations are the geometric form of the Lagrange equations.

So, let the external fields be known: the potential energy functions $U=U (q^1, q^2,…, q^s, t)$, the potential momentum $P_{\alpha}=P_{\alpha}(q^1,q^2,…,q^s,t)$ and the metric tensor of the configuration space $\mu_{\alpha\beta}=\mu_{\alpha\beta}(q^1,q^2,…,q^s,t)$. We will also assume that the total energy of the system is known as a function of time $E(t)$. It is required to establish an equation of the form
\[ \frac{d^2q^{\alpha}}{dq^2}=f^{\alpha}\left(q^{\beta},\frac{dq^{\gamma}}{dq},t\right)      \]
or equivalent to it.

To do this, it is necessary to replace all velocities $\dot{q}^{\alpha}$ and accelerations $\ddot{q}^{\alpha}$ in the Lagrange equations \eqref{voy1.6}
respectively with the components of the tangent vector $\tau^\alpha=dq^{\alpha}/dq$ and the components of the vector $d{\tau}^{\alpha}/dq$. Let's do this: replace the differentiation variable - time $t$ in the generalized velocities and accelerations with the length of the curve $q$ according to \eqref{voy1.8}. We get
\begin{equation}\label{voy2.1}
	\dot{q}^\beta=\frac{dq}{dt}\frac{dq^\beta}{dq}=\sqrt{2T}\,\,\tau^{\beta}\,,
\end{equation}
	\[\ddot{q}^\beta=\frac{d}{dt}\left(\frac{dq^\beta}{dt}\right)=\sqrt{2T}\,\frac{d}{dq}\left(\sqrt{2T}\frac{dq^\beta}{dq}\right)=2T\,\frac{d\tau^\beta}{dq}+\frac{dT}{dq}\,\tau^\beta=\]
	\begin{equation}\label{voy2.2}
	=2T\,\,\frac{d\tau^\beta}{dq}+\frac{1}{\sqrt{2T}}\,\,\frac{dT}{dt}\,\,\tau^\beta\,.
\end{equation}
From the equality \eqref{voy2.2} it is clear that only it and the expression \eqref{voy2.1} are not enough to derive the trajectory equation. It is also necessary to know how the kinetic energy changes over time.

\subsubsection{Rate of change of kinetic energy}

Let us find out how the rate of change of the kinetic energy $T$ of a natural system is related to the change of external fields. The theorem (on the change of the kinetic energy of a natural system) is valid:

\textit{The change of the kinetic energy T of a natural system as a function of time is related to the external fields acting on it by the following equality:}
\begin{equation}\label{voy3.1}
	\frac{dT}{dt}=\left(-\frac{\partial U}{\partial q^{\gamma}}-\frac{\partial P_{\gamma}}{\partial t}\right)\dot{q}^{\gamma}-\frac{1}{2}\frac{\partial \mu_{\gamma\alpha}}{\partial t}\,\,\dot{q}^{\alpha}\dot{q}^{\gamma}\,.
\end{equation}

Proof. Let us find the total derivative of the kinetic energy \eqref{voy1.7} with respect to time. We obtain that
\[ \frac{dT}{dt}= \frac{1}{2}\frac{\partial \mu_{\alpha\beta}}{\partial t}\dot{q}^{\alpha}\dot{q}^{\beta}+ \mu_{\alpha\beta}\,\dot{q}^{\alpha} \ddot{q}^{\beta}\,.\]
Let us substitute into this equality the expression \eqref{voy1.4} and the equation of motion \eqref{voy1.6} and group the terms by the powers of the velocities
\[	\frac{dT}{dt}=\left(-\frac{\partial U}{\partial q^{\gamma}}-\frac{\partial P_{\gamma}}{\partial t}\right)\dot{q}^{\gamma}+\left(\frac{\partial P_{\alpha}}{\partial q^{\gamma}}-\frac{\partial P_{\gamma}}{\partial q^{\alpha}}-\frac{1}{2}\frac{\partial \mu_{\gamma\alpha}}{\partial t}\right)\dot{q}^{\alpha}\dot{q}^{\gamma}+\]
	\begin{equation}\label{voy3.2}
	+\frac{1}{2}\left(\frac{2\partial\mu_{\alpha\beta}}{\partial q^{\gamma}}-\frac{\partial\mu_{\gamma\beta}}{\partial q^{\alpha}}-\frac{\partial\mu_{\gamma\alpha}}{\partial q^{\beta}}\right)\dot{q}^{\alpha}\dot{q}^{\beta}\dot{q}^{\gamma}\,.
\end{equation}
In the second bracket \eqref{voy3.2} the derivatives of the potential momentum are antisymmetric with respect to the indices $\alpha$ and $\gamma$ and yield zero when multiplied by the symmetric form $\dot{q}^{\alpha}\dot{q}^{\gamma}$. Therefore, they can be excluded from this bracket. The bracket of the third term is transformed as follows:
\[\frac{1}{2}\left(\frac{2\partial\mu_{\alpha\beta}}{\partial q^{\gamma}}-\frac{\partial\mu_{\gamma\beta}}{\partial q^{\alpha}}-\frac{\partial\mu_{\gamma\alpha}}{\partial q^{\beta}}\right)\dot{q}^{\alpha}\dot{q}^{\beta}\dot{q}^{\gamma}=\frac{1}{2}\left(\frac{\partial\mu_{\alpha\beta}}{\partial q^{\gamma}}-\frac{\partial\mu_{\gamma\beta}}{\partial q^{\alpha}}+\frac{\partial\mu_{\alpha\beta}}{\partial q^{\gamma}}-\frac{\partial\mu_{\gamma\alpha}}{\partial q^{\beta}}\right)\dot{q}^{\alpha}\dot{q}^{\beta}\dot{q}^{\gamma}\,.\]
From this it is clear that the first difference of the derivatives is antisymmetric in the indices $\alpha$ and $\gamma$, and the second difference of the derivatives is antisymmetric in the indices $\beta$ and $\gamma$. Therefore, the result of multiplying the bracket by the symmetric form $\dot{q}^{\alpha}\dot{q}^{\beta}\dot{q}^{\gamma}$ is zero. Taking these circumstances into account, the right-hand side of \eqref{voy3.2} is simplified to the right-hand side of \eqref{voy3.1} $\blacksquare$ . The second term in \eqref{voy3.1} could be expected in advance, since with an increase in the moment of inertia of a rotating body, its kinetic energy decreases.

The theorem on the change in kinetic energy of a natural system \eqref{voy3.1} can be rewritten in an equivalent geometric form by substituting the equalities \eqref{voy2.1} into \eqref{voy3.1}. Then we obtain
\begin{equation}\label{voy3.3}
	\frac{dT}{dt}=\sqrt{2T}\,\left(-\frac{\partial U}{\partial q^{\gamma}}-\frac{\partial P_{\gamma}}{\partial t}\right)\tau^{\gamma}-T\,\frac{\partial \mu_{\gamma\alpha}}{\partial t}\,\,\tau^{\alpha}\tau^{\gamma}\,.
\end{equation}

\subsubsection{Lagrange equations in geometric form}

Let us derive the equations of the trajectory. Substituting equality \eqref{voy3.3} into \eqref{voy2.2}, we obtain that the generalized accelerations depend on the tangent vector and the vector $d\tau^\beta/dq$ as follows
\begin{equation}\label{voy4.1}
	\ddot{q}^\beta=2T\,\,\frac{d\tau^\beta}{dq}+\left(-\frac{\partial U}{\partial q^{\gamma}}-\frac{\partial P_{\gamma}}{\partial t}\right)\,\tau^\beta \tau^{\gamma}-\frac{\sqrt{2T}}{2}\,\frac{\partial \mu_{\gamma\alpha}}{\partial t}\,\,\tau^{\alpha} \tau^{\beta}\tau^{\gamma}\,\,.
\end{equation}

Now we substitute the equalities \eqref{voy2.1} and \eqref{voy4.1} into the equation of motion \eqref{voy1.6}, where we take into account that $T=E-U$ and leave on the left side of the resulting equality only the term proportional to $d\tau^{\beta}/dq$. After this, we group all the terms on the right side by powers of the tangent vector. As a result, we obtain
\[	2(E-U)\mu_{\gamma\beta}\frac{d\tau^{\beta}}{dq}=-\frac{\partial U}{\partial q^{\gamma}}-\frac{\partial P_{\gamma}}{\partial t}+\sqrt{2(E-U)}\left(\frac{\partial P_{\alpha}}{\partial q^{\gamma}}-\frac{\partial P_{\gamma}}{\partial q^{\alpha}}-\frac{\partial \mu_{\gamma\alpha}}{\partial t}\right)\,\tau^{\alpha}+\]
	\[	+\left[(E-U)\left(\frac{\partial\mu_{\alpha\beta}}{\partial q^{\gamma}}-\frac{\partial\mu_{\gamma\beta}}{\partial q^{\alpha}}-\frac{\partial\mu_{\gamma\alpha}}{\partial q^{\beta}}\right)+\mu_{\gamma\beta}\left(\frac{\partial U}{\partial q^{\alpha}}+\frac{\partial P_{\alpha}}{\partial t}\right)\right]\,\tau^{\alpha} \tau^{\beta}+\]
	\begin{equation}\label{voy4.2}
	+\frac{\sqrt{2(E-U)}}{2}\mu_{\gamma\beta}\frac{\partial \mu_{\epsilon\alpha}}{\partial t}\,\tau^{\alpha} \tau^{\beta}\tau^{\epsilon}\,.
\end{equation}

These equalities are the sought general equations of the trajectory of a natural system in the configuration space with the metric tensor $\mu_{\alpha \beta}$. The number of these equations is equal to the number of degrees of freedom $s$ of the natural system.

\vspace{2\baselineskip}

\begin{flushleft}
{\bf{Discussion}}
\end{flushleft}

It is easy to see that for the case of a material point ($\mu_{\alpha \beta}=m\delta_{\alpha\beta}$) in a constant non-magnetic field (${\partial P_{\gamma}}/{\partial t}=0$, ${\partial P_{\alpha}}/{\partial q^{\gamma}}=0$) the obtained trajectory equations coincide with \cite[problem to item 44]{1}, \cite[item 2.7.]{2}. As another check of \eqref{voy4.2}, let us multiply both sides of this equation by the vector $\tau^\gamma$. After multiplication and performing obvious transformations, the left side of the resulting equality takes the form
\[2(E-U)\mu_{\gamma\beta}\tau^\gamma\frac{d\tau^{\beta}}{dq}=(E-U)\left[\frac{\left(d\mu_{\gamma\beta}\tau^\gamma\tau^\beta\right)}{dq}-\tau^\gamma\tau^\beta\frac{d\mu_{\gamma\beta}}{dq}\right]=-(E-U)\tau^\gamma\tau^\beta\frac{d\mu_{\gamma\beta}}{dq}=\]
\[=--(E-U)\left(\frac{\partial \mu_{\gamma\beta}}{\partial q^\alpha}\tau^\alpha+\frac{1}{\sqrt{2(E-U)}}\frac{\partial \mu_{\gamma\beta}}{\partial t}\right)\tau^\gamma\tau^\beta=
-(E-U)\frac{\partial \mu_{\gamma\beta}}{\partial q^\alpha}\tau^\alpha\tau^\beta\tau^\gamma-\]
\[-\frac{\sqrt{2(E-U)}}{2}\frac{\partial \mu_{\gamma\beta}}{\partial t}\tau^\beta\tau^\gamma\]
The right side is equal to
\[	-\frac{\partial U}{\partial q^{\gamma}}\tau^\gamma-\frac{\partial P_{\gamma}}{\partial t}\tau^\gamma+\sqrt{2(E-U)}\left(\frac{\partial P_{\alpha}}{\partial q^{\gamma}}-\frac{\partial P_{\gamma}}{\partial q^{\alpha}}-\frac{\partial \mu_{\gamma\alpha}}{\partial t}\right)\,\tau^{\alpha}\tau^\gamma+\]
	\[	+\left[(E-U)\left(\frac{\partial\mu_{\alpha\beta}}{\partial q^{\gamma}}-\frac{\partial\mu_{\gamma\beta}}{\partial q^{\alpha}}-\frac{\partial\mu_{\gamma\alpha}}{\partial q^{\beta}}\right)+\mu_{\gamma\beta}\left(\frac{\partial U}{\partial q^{\alpha}}+\frac{\partial P_{\alpha}}{\partial t}\right)\right]\,\tau^{\alpha} \tau^{\beta}\tau^\gamma+\]
	\[	+\frac{\sqrt{2(E-U)}}{2}\,\mu_{\gamma\beta}\,\frac{\partial \mu_{\epsilon\alpha}}{\partial t}\,\tau^{\alpha} \tau^{\beta}\tau^{\epsilon}\tau^\gamma=-\frac{\partial U}{\partial q^{\gamma}}\tau^\gamma-\frac{\partial P_{\gamma}}{\partial t}\tau^\gamma-\sqrt{2(E-U)}\,\,\frac{\partial \mu_{\gamma\alpha}}{\partial t}\,\tau^{\alpha}\tau^\gamma+\]
\[+(E-U)\left(\frac{\partial\mu_{\alpha\beta}}{\partial q^{\gamma}}-\frac{\partial\mu_{\gamma\beta}}{\partial q^{\alpha}}-\frac{\partial\mu_{\gamma\alpha}}{\partial q^{\beta}}\right)\tau^{\alpha} \tau^{\beta}\tau^\gamma+\left(\frac{\partial U}{\partial q^{\alpha}}+\frac{\partial P_{\alpha}}{\partial t}\right)\tau^{\alpha}+ \]
\[+\frac{\sqrt{2(E-U)}}{2}\frac{\partial \mu_{\epsilon\alpha}}{\partial t}\,\tau^{\alpha}\tau^{\epsilon}=-(E-U)\frac{\partial \mu_{\gamma\beta}}{\partial q^\alpha}\tau^\alpha\tau^\beta\tau^\gamma-\frac{\sqrt{2(E-U)}}{2}\frac{\partial \mu_{\gamma\beta}}{\partial t}\tau^\beta\tau^\gamma\]
We obtain the identity as it should be. In other words, equations \eqref{voy4.2} are mathematically consistent.

The geometric properties of a trajectory in a configuration space are, according to \eqref{voy4.2}, a consequence of variable external influences. These equations are not purely geometric, since the fields $U$, $P_\alpha$, $\mu_{\alpha\beta}$ may depend on time. Such equations are called \textit{nonautonomous} \cite{12}. Despite the fact that these equations implicitly include time, this does not exclude them from the field of differential geometry. On the contrary, such time dependence complements the geometric properties of natural systems with various kinds of nonlinear effects. As an example of the existence of a time variable in geometric equations, we can mention the equation of a geodesic in a curved space-time, which implicitly includes time \cite{11}.

The equations for the trajectory \eqref{voy4.2} are complex. It is very likely that complex methods of differential geometry and analysis will be required for the analytical solution of these equations. The solution of the differential equation \eqref{voy4.2}, describing the change in the tangent vector, depends on the initial conditions, external fields and the properties of the natural system itself (the metric tensor of the configuration space $\mu_{\alpha\beta}$). The initial conditions are the initial tangent vector $\tau^{\alpha}_0$, the initial coordinates $q^{\alpha}_0$ and the initial time $t_0$. The numerical method that can be used to find the solution is the popular and accurate Runge-Kutta method of the 4th order (RK4).

If the external fields change sufficiently smoothly and slowly in space and time, then we can expect that the solution exists and is unique in some neighborhood of the initial position of the system. If this condition is not met (that is, singularities or features arise), then in such cases special methods may be required to ensure the stability and accuracy of the solution.

\begin{flushleft}
{\bf{Conclusion}}
\end{flushleft}

In this paper, the general Lagrangian formulation \eqref{voy3.1} and the geometric trajectory formulation \eqref{voy3.3} of the well-known theorem on the change of kinetic energy (e.g., \cite{8}, \cite{9}) for natural systems were clarified. By applying it to the Lagrange equations, the trajectory equations of the natural system \eqref{voy4.2} moving under the action of variable external fields were derived. The trajectory equations together with the equality \eqref{voy1.8} form a new, trajectory method for solving problems of dynamics. Thus, the geometric view of non-stationary mechanical processes is substantiated.

The question of how the trajectory equations are related to the principle of extremal action in Jacobi form remains open. This problem is the topic of one of the following papers.

\begin{flushleft}
{\bf{Information after article in English}}
\end{flushleft}
Voytik V.V. Trajectory equations for a non-conservative natural system. \textit{Vestnik 
Tomskogo gosudarstvennogo universiteta. Matematika i mekhanika} [Tomsk State University Journal of Mathematics and Mechanics].  

\textit{Abstract.} In practice, it is often necessary to know the trajectory of motion of natural mechanical systems. At present, the trajectory equations in the configuration space are well known only for some conservative systems, which makes it important to derive equations for systems in non-stationary external fields. In this paper, we prove a theorem on the change in kinetic energy, which states that the rate of its change depends both on external forces and on the rate of change of the metric tensor. This theorem can be geometrically expressed as a combination of products of forces and changes in the metric tensor with tangent vectors. Generalized velocities and accelerations are similarly described in terms of tangent vectors and their derivatives along the trajectory. Substituting these expressions into the Lagrange equations leads to trajectory equations corresponding to the degrees of freedom of the system. The left-hand side contains the covariant derivative of the tangent vector, and the right-hand side includes a cubic polynomial of tangent vectors. These equations represent the geometric form of the Lagrange equations, which can be solved numerically using the 4th order Runge-Kutta method. Together with the trajectory parameterization, they create a trajectory method for solving dynamics problems.

\end {document}